\documentstyle[11pt,paspconf]{article}
\newcommand{\teff}{$T_{\rm eff}$}
\newcommand{\vsini}{$v \sin i$}

\begin{document}

\title{Radiative Levitation in Hot Horizontal Branch Stars}
\author{W. Landsman}
\affil{Raytheon ITSS, Code 681, NASA/GSFC, Greenbelt, MD 20770}

\begin{abstract}
There is now considerable evidence that  horizontal branch (HB) stars hotter
than about 11,500~K experience an enormous enhancement of their photospheric
iron abundance due to radiative levitation.   In globular clusters, the photospheric iron abundance
can reach values of [Fe/H] $\sim +0.3$, or up to two orders of magnitude higher
than the cluster iron abundance.    Model atmospheres which take into
account the iron overabundance are needed for understanding the appearance of
the HB in globular cluster color-magnitude diagrams (CMDs),  for the derivation of 
accurate luminosities, gravities and masses, and for the ultraviolet spectral
synthesis of old, metal-poor stellar populations.    
\end{abstract}


\section{Introduction}
Grundahl et al.\ (1999) recently suggested that radiative levitation of heavy
elements to supersolar abundances occurs for globular cluster HB stars hotter
than about $11,\!500$~K.    This suggestion was based on the following
evidence:

\begin{itemize}

\item  Grundahl et al.\ obtained Str\"{o}mgren  $u$, $u-y$  CMDs of 14 globular
clusters and found evidence for a ubiquitous ``jump'' at about \teff\ =
$11,\!500~{\rm K}$, in the sense that stars hotter than this temperature  are
about 0.25 mag brighter in Str\"{o}mgren $u$ than predicted by HB models.  
This jump is {\em not} present in ultraviolet CMDs of globular clusters
obtained with HST or the Ultraviolet Imaging Telescope (UIT), which suggests
that it is due to an atmosphere effect (causing a redistribution in the flux),
rather than to a change in the bolometric luminosity.    (The presence of a
luminosity jump is difficult to discern at wavelengths longer than
Str\"{o}mgren $u$ because the HB becomes ``vertical" in the CMD, and the
effects of changes in luminosity cannot be distinguished from changes in
temperature.)

\item As summarized by Moehler (1999), the derived gravities of HB
stars with $11,\!500~{\rm K} < T_{\rm eff} < 20,\!000$~K are consistently
found to be about 0.2 dex lower than predicted by canonical HB models.    
Grundahl et al.\ performed simple experiments with Kurucz model atmospheres
which suggested that both the gravity anomaly and the brightening in 
Str\"{o}mgren $u$, could be explained if the HB photosphere had a supersolar 
metallicity, rather than the metallicity of the cluster.    

\item Hot HB stars in globular clusters are observed to show strong depletions
of helium (e.g. Moehler et al.\ 1997).
The early theoretical work of Michaud et al. (1983) suggested that if the
HB atmosphere is stable enough to allow for the gravitational settling of
helium, then overabundances of heavy elements by factors of $10^3 - 10^4$ might
be expected due to radiative levitation.

\item Helium-depleted field HB stars hotter than $11,\!500~{\rm K}$ show
unusual abundance patterns, and have higher iron abundances than observed in
cooler HB stars.   In particular, in the well-studied field HB star Feige 86
(Castelli et al.\ 1997), the elements lighter than sulfur are depleted (with
the exception of phosphorus, with [P/H] = +1.8), while iron-peak elements
are slightly supersolar  ([Fe/H] = +0.4), and the heavy metals are strongly
overabundant (e.g.\ [Au/H] = +4.0).    Glaspey et al.\ (1989) reported
an overabundance of iron by a factor of 50 (and a helium depletion) 
in a hot HB star in NGC~6752 (CL~1083) with  $T_{\rm eff} = 16,\!000$~K.
\end{itemize}

With the exception of the single star in NGC~6752 studied (at low S/N) by
Glaspey et al., all the evidence for an iron enhancement in globular cluster
HB stars discussed by Grundahl was indirect.    However, independent of
Grundahl et al.\ work,  Behr et al.\ (1999) were using the Keck HIRES echelle
spectrograph to study abundances in 13 hot HB stars in M13.     Their results
provide striking direct evidence for radiative levitation in globular cluster
hot HB stars, and for abundance patterns similar to those observed in Feige
86. The iron abundances in the M13 HB stars hotter than $11,\!500$~K are about
$+2.0$ dex  higher than in stars cooler than the jump temperature.     Phosphorus is also enhanced
([P/H] = +1.0), but the magnesium abundances ([Mg/H] $\sim -1.5$) show no
change across the jump temperature. Subsequently, Moehler et al.\ (1999) used
medium ($\sim 2.6$ \AA) resolution spectroscopy to show the presence of a similar jump
to supersolar iron abundances  (with Mg remaining at the cluster abundance) in
NGC~6752. In addition, Moehler et al. were able to explicitly show that most
-- though not all -- of the discrepancy between the derived gravities  and
canonical models could be removed if the  Balmer lines were analyzed using
appropriately metal-rich atmospheres.      We hope to further explore the
appropriate model atmospheres for hot HB stars using our Cycle 8 HST program
(8256) to obtain STIS ultraviolet spectra of nine HB stars in NGC~6752
spanning the temperature range of $10,\!000~{\rm K} < T_{\rm eff} <
24,\!000$~K. 

The sample of stars with abundances determined by Behr et al.\ or  Moehler et
al.\ does not include any stars hotter than 20,000~K.   Peculiar abundance
patterns are known to exist in the sdB stars (e.g.\ Lamontagne et al.\ 1987),
and pulsation studies indicate the presence of radiative levitation of iron
within the envelope (Charpinet et al.\ 1997).     However, the photometric
discrepancy with canonical models in Str\"{o}mgren  $u$ discussed by Grundahl
et al.\ decreases for  \teff $>$ 20,000 K, and the field sdB stars do not
show the strong {\em over}abundances seen in cooler (and lower gravity) HB
stars, such as Feige 86. Thus, we suggest that the most dramatic effects of
radiative levitation occur in the temperature range $11,\!500~{\rm K} < T_{\rm
eff} < 20,\!000$~K. 

The presence of strong abundance anomalies in hot HB stars complicates the
derivation of accurate luminosities, gravities, and masses, and the analysis of
the integrated ultraviolet spectra of old, metal-poor stellar populations.  But
these abundance anomalies also mean that the HB of globular clusters will
likely provide a superb laboratory for studying radiative levitation and 
diffusion processes in the outer atmospheres of hot stars. The HB stars in a
globular cluster have known initial abundances, and provide a populous sample
for studying the effects of temperature, initial metallicity, and rotation on
the photospheric abundances resulting from radiative levitation.   (The
uniformity of the Str\"{o}mgren $u$ jump indicates that the effects of
radiative levitation must be rapid compared to the HB lifetime  of about $10^8$
yr.)

In analogy to the HgMn stars (e.g.\ Leckrone et al.\ 1999), the rough
abundance pattern observed in hot HB stars can be understood  as being due to
saturation of radiative forces in the more abundant lighter elements.   
However other effects reported by Behr et al.\ and Moehler et al.\ have no
ready explanation, including the increased  helium depletion  with
increasing \teff, and in particular, the   abruptness of the transition to
supersolar iron abundances at 11,500 K.     An important clue might be
provided by the observation by Behr et al. of a low rotation (\vsini $< 6$ km
s$^{-1}$)  in stars hotter than the jump, since rotationally-induced turbulence
can inhibit diffusion processes.    But this only moves the problem one level
deeper, since the origin of the abrupt change in rotational
velocities would still be unknown.

\section{Implications}
What are the implications of the discovery of radiative levitation for the
outstanding problems in the studies of globular cluster HB stars? First, the
discovery of  radiative levitation has no direct implications for the age or
distance scale of globular clusters, since hot HB stars have not been used as
absolute calibrators.    Similarly, the discovery of  radiative levitation in
hot HB stars does not answer the question of {\em why} some HB stars are hot,
or, more generally, on the origin of the HB morphology (e.g. the second
parameter problem).

Another outstanding problem in HB studies is the origin of ``gaps" in the
temperature distribution of HB stars, and in this case, radiative levitation
might have a contributing role for any gap located near $11,\!500$~K (such as the
G1 gap discussed by Ferraro et al.\ 1998). In certain bandpasses, the sudden
change in photospheric abundances near this temperature might shift the
positions of stars along the HB.    For example, in a $y$, $u-y$ diagram, the
brightening in Str\"{o}mgren $u$ with the onset of radiative levitation
 could induce a gap in the HB distribution at $11,\!500$~K.

Although the dominant implications of radiative levitation are for the HB
stellar atmosphere, in principle, the changes in the radial chemical profile 
could alter the bolometric luminosities and lifetimes computed in HB interior
models.  This effect is likely to be small, although Seaton (1997) does warn
that the surface abundance changes patterns in the HgMn stars are probably a
manifestation of   radiative diffusion processes deep in the stellar
envelope, which (through the modified opacities) can alter the stellar
structure. 

Hot HB stars are one of the few potential ultraviolet sources in an old stellar
population, and thus the discovery of radiative levitation has significant
implications for the ultraviolet spectral synthesis of old, metal-poor
systems.   The implications are less for a metal-rich population for two
reasons: first, the metallicity enhancement due to radiative levitation is less
pronounced for metal-rich stars, and second, fewer hot HB stars are expected in
a metal-rich system, since the HB is expected to bifurcate into either very hot
($> 20,000$~K)  or cool HB stars (Dorman, Rood, \& O'Connell 1995).    Spectral
synthesis models of the ultraviolet upturn in elliptical galaxies computed in
either the metal-poor or the mixed metallicity scenarios discussed
in Yi et al.\ (1999), should probably be performed using metal-rich atmospheres
for metal-poor HB stars with  $T_{\rm eff} > 11,\!500$~K.    In particular, the
use of metal-rich model atmospheres might help suppress the model flux in the
1800 -- 2500 \AA\ spectral region, and improve the agreement of the
metal-poor models shown in Yi et al.\ with the observed spectra of ellipticals.

Finally, the empirical finding of Behr et al.\ (1999) and Moehler et al.\ (1999)
that Mg abundances are unaltered by diffusion processes suggests that Mg is the
most reliable abundance indicator for field hot HB stars.  For example, while
most the of the abundances derived in Feige 86 by Castelli et al.\ (1997) have
been modified by diffusion processes, the magnesium abundance ([Mg/H] = --0.64)
probably provides a good measure of the stellar metallicity.

\acknowledgments
I thank my collaborators on this topic, including M. Catelan, F. Grundahl, T.
Lanz, S. Moehler, C. Proffitt, and A. Sweigart

\end{document}